\documentstyle[12pt,epsfig]{article}
\begin{document}
\baselineskip 22pt

\vspace{3.0cm}

\begin{center}
{\Large \bf On the Gauged Non-compact Spin System}\\
\vspace{.5cm}

Sung-Soo Kim\footnote{Electronic-mail: billy@newton.skku.ac.kr}
and Phillial Oh\footnote{Present Address:
Center for Theoretical Physics, M.I.T., Cambridge, MA 02139-4307}\\
{\it Department of Physics, Sung Kyun Kwan University,
Suwon 440-746,  Korea }\\
\end{center}

\vspace{0.5cm}

\begin{center} {\bf Abstract} \\ 
\end{center} 
We examine classical and
quantum aspects of the planar non-compact spin system coupled with
Chern-Simons gauge field in the presence of background charge. We first
define our classical spin system as non-relativistic non-linear sigma
model in which the order parameter spin takes value in the non-compact
manifold ${\cal M}=SU(1,1)/U(1)$.  Although the naive model does not allow
any finite energy self dual solitons, it is shown that the gauged system
admits static Bogomol'nyi solitons with finite energy whose rotationally 
symmetric soliton solutions are analyzed in detail.  
We also discuss the large spin limit in which the self-dual equation 
reduces to the well-known gauged nonlinear Schr\"odinger model or 
Abelian Higgs model, depending on the choice of the background charge term. 
Then, we perform quantization of the model. We find that the spin algebra 
satisfies anomalous commutation relations, and the system is a field 
theoretic realization of the anyons.  

\hspace{0.3cm}


\baselineskip 18pt

\thispagestyle{empty}
\pagebreak

\baselineskip 18pt
\section{Introduction}

Recently, a nonrelativistic nonlinear sigma model (NLSM) defined on the 
target space of coadjoint orbits $G/H$ was proposed, whose 
Euler-Lagrange equation of motion yields the generalized continuous 
Heisenberg  ferromagnet~\cite{ohpark1}. In this system, 
the dynamical variable is spin defined on coadjoint 
orbit  whose Poisson bracket satisfies the classical ${\cal G}$ algebra. 
When the target spaces of coadjoint orbits are given by Hermitian 
symmetric spaces (HSS)~\cite{Helgason} which are symmetric spaces 
equipped with complex structure, the generalized spin system becomes  
completely integrable in 1+1 dimension. It was also discovered that 
incorporation of the Chern-Simons (CS) gauge field in 2+1 dimension
on the same target space produces  a class of self-dual field theories. 
They admit Bogomol'nyi type self-dual equations whose energy is 
saturated by the topological charge~\cite{ohpark2}. A detailed numerical
investigation in the compact $SU(2)$ case \cite{ohkim}
showed a rich structure of rotationally symmetric
self-dual CS solitons with finite energies.

Since the self-dual CS
solitons attracted a upsurge of recent theoretical interest \cite{dunn},
it is worthwhile to give a more detailed investigation
to the gauged spin system.
In this paper, we carry out this in the case of
CS gauged non-compact symmetry. We consider the 
simplest non-compact $G=SU(1,1)$ with the target space of
the hyperboloid ${\cal M}=SU(1,1)/U(1)$ which is a 
well-known example of the non-compact
HSS. The motivations for non-compact symmetry are two-fold.
On the classical aspect, we are interested in 
the Bogomol'nyi self-dual solitons with finite energy.
First of all, the non-compact symmetry in
nonlinear sigma model was considered 
in a wide context before, in relation with dynamical 
generation of gauge bosons in supergravity theory~\cite{davis}, 
integrable non-linear Schr\"odinger equation in 1+1 dimension 
~\cite{kund}, Ernest equation in 2+1 dimension~\cite{grus},
and also in the context of condensed 
matter physics such as integer quantum Hall 
effect~\cite{prui} and magnetic spin system \cite{makh}.
The static self-dual Bogomol'nyi 
solitons with non-compact symmetry
were studied only recently in ~\cite{kori} where it was shown that
relativistic NLSM on ${\cal M}$ coupled 
with the CS gauge  field admits rotationally symmetric
soliton solutions.
On the non-relativistic side, the result of 
Ref. \cite{ohpark2} on self-dual
CS theories in the arbitrary
HSS implies that the non-compact spin system will 
admit the self-dual equations with Bogomol'nyi bound.
 But this does not guarantee the existence of 
finite positive semi-definite
energy solitons. One of our main purpose is to survey the existence
of static self-dual Bogomol'nyi solitons with finite energy
in the  gauged non-compact spin system.

The  non-compact spin variables are defined on the target space
${\cal M}$. In contrast to  the 
compact $SU(2)$ case on $S^2$ which is
a typical example of circumventing the well-known no-go theorem
in static solitons~\cite{hoba}, and is 
supported by the finite energy topological lump solutions
characterized by the integer winding, $\pi_2(S^2)={\bf{Z}}$ \cite{bela},
the ungauged non-compact spin system does not allow any static solitons 
with an integer winding because of $\pi_2({\cal M})=0$.  
Besides, the self-dual equations have only
singular solutions in which both the energy and 
topological charge diverge
(see discussion following (\ref{sing})).
Coupling with other matter field 
improves the situation in that there exist 
exact solutions characterized by a finite topological charge, 
but they are singular and not localized soliton solutions~\cite{leo}. 
Incorporation of the gauge field drastically changes the
situation.   A detailed investigation of 
the gauged self-dual equations and their numerical
solutions indeed reveals  that the gauged non-compact spin 
system admits self-dual CS solitons with finite energy. 

Another aspect of interest is that the system permits classical
Holstein-Primakoff transformation ~\cite{oh966} 
($\psi\rightarrow\vartheta$) which changes
the symplectic one-form on ${\cal M} (\sim 
\frac{\psi^*\dot\psi-\dot{\psi}^* \psi}{1-\vert\psi\vert^2})$ into
the canonical structure ($\sim \vartheta^*\dot\vartheta)$. 
This transformation has an one-parameter dependence on the 
total spin and we find that
in the large spin limit, the self-dual equations reduce to
the well-known gauged nonlinear Schr\"odinger model of Jackiw and
Pi~\cite{jack-pi}, or Abelian Higgs model~\cite{wein}, 
depending on the choice of the background charge term.

The second is quantum aspect, especially concerned with realization
 of anyons ~\cite{welc}.
It is well-known that the CS field produces the
long-range interaction which is responsible for the existence
of anyon. We find that  the quantization of the gauged non-compact 
spin system also leads to the nonrelativistic field theory describing 
anyons, and in the large spin limit it again reduces to the gauged 
nonlinear  Schr\"odinger model.
Of course, these features are not something which belongs only to 
non-compact symmetry. The same analysis performed in this paper also
indicates that the compact version of the model~\cite{ohkim} also 
exhibits the anyon characteristic. But it is interesting to 
have these properties in the non-compact case also.  

The paper is organized as follows: In Section 2, we first describe the 
ungauged non-compact spin system and set the framework for
further discussion. Then, we couple with the CS  gauge field 
and study the self-dual solitons. We also discuss large spin limit. 
In Section 3, we quantize the gauged non-compact spin system and derive
the anomalous commutation relations.
Section 4 is the conclusion and discussion.

\section{Noncompact Spin and CS Solitons}
We first recall that the action principle for the (2+1)-dimensional
generalized Heisenberg ferromagnet model defined on the coadjoint
orbit  $G/H$ is given by \cite{ohpark1}
\begin{equation}
S= \int dt d^2x  \mbox{ tr }\left[2Kg^{-1}\dot{g}-
\partial_{i}(gK g^{-1})\partial_{i} (gK g^{-1})\right],
\label{preact}
\end{equation}
where $g$ is a map $g:R^{2+1} \rightarrow G$ for the group $SU(1,1)$,
and $K=i\frac{J}{2}~\mbox{diag}(1,-1)$. $J$ is related with the
representation of $SU(1,1)$ group \cite{pere}.
Let us express the element $g$ of $SU(1,1)$
by the kets $(\vert \Psi_1\rangle, \vert \Psi_2\rangle)$
with $ \vert \Psi_p \rangle=(\Psi_p^{(1)},\Psi_p^{(2)})^T~(p=1,2)$,
and denote  the inner product between the ket and
its bra $\langle\bar\Psi_q\vert= ( \Psi^*_{q(1)}, \Psi^*_{q(2)})$
via  $\langle\bar \Psi_q\vert \Psi_p \rangle
\equiv\sum_{i=1,2}\Psi^*_{q(i)}\Psi_p^{(i)}.$
We introduce the metric $M^{ij}=M_{ij}=
\mbox{diag}(1, -1)\;~(i,j=1,2)$
with respect to which the raising and lowering of the indices
of the components of the bra and ket are performed:
$\Psi_p^{*(i)}=M^{ij}\Psi^*_{p(j)}$ and
$\Psi_{p(i)}=M_{ij}\Psi_p^{(j)}$. Then, the inner product can be
written as $\langle\bar \Psi_q\vert \Psi_p\rangle=\sum_{i=1,2}
\Psi^{*(i)}_qM_{ij}\Psi_p^{(j)}$.
The $SU(1,1)$ condition, $g^\dagger Mg=M$ gives
\begin{equation}
\langle \bar \Psi_p\vert \Psi_q\rangle=M_{pq},~~
\mbox{det}(\vert \Psi_1\rangle, \vert \Psi_2\rangle)=1.
\label{cond}
\end{equation}  

By making use of the second equation of (\ref{cond}),
$\vert \Psi_2\rangle$ can be eliminated and the non-compact spin
variable $Q$ can be expressed by
\begin{equation}
Q\equiv
gKg^{-1}=iJ\left(\vert\Psi_1\rangle\langle
\bar\Psi_1\vert-\frac{1}{2}I \right)=
Q^at^b\eta_{ab},\label{isosp1}
\end{equation}
where $t^a$'s are given by the Pauli matrices;
$t^1=-\frac{1}{2}\sigma_2,~ t^2=-\frac{1}{2}\sigma_1,~
t^3=i\frac{1}{2}\sigma_3$.
They satisfy the Lie algebra:
$[t^1, t^2]=-t^3,~ [t^2, t^3]=t^1,~[t^3, t^1]=t^2$ and
$\mbox{tr}(t^at^b)=-\frac{1}{2}\eta^{ab},~ \eta^{ab}=
\mbox{diag}(-1,-1,1)$,
with the constraint $\langle \bar \Psi_1\vert \Psi_1\rangle=1$.
This constraint can be solved explicitly in terms of the
complex projective coordinate defined by
$\psi^*=\Psi_1^{(2)}/\Psi_1^{(1)}~(\Psi_1^{(1)}\neq 0)$
with a real gauge condition;
\begin{equation}
\chi=\frac{1}{2}( \Psi^{*(1)}_1 -\Psi^{(1)}_1) = 0\label{gauge1}.
\end{equation}
Then, the solution to the above constraint is given by
\begin{equation}
\Psi^{*(1)}_1=\Psi^{(1)}_1=\frac{1}{\sqrt{1-\vert \psi\vert^2}}.
\label{sol1}
\end{equation}
Putting everything together into (\ref{preact}),
we obtain the following reduced Lagrangian:
\begin{equation}
{\cal L}_\psi=iJ \frac{{\psi}^* \dot{\psi}-\dot{\psi^*}\psi}
{1-\vert\psi\vert^2}-J^2
\frac{2\partial_i\psi^*\partial_i\psi}
{(1-\vert\psi\vert^2)^2},
\label{xilag} 
\end{equation}
and  the non-compact spin from (\ref{isosp1}) follows that
\begin{equation}
Q^1=J\frac{\psi+\psi^*}{1-\vert\psi\vert^2},~~
Q^2=iJ\frac{\psi^*-\psi}{1-\vert\psi\vert^2},~~
Q^3=J\frac{1+\vert\psi\vert^2}{1-\vert\psi\vert^2}.   
\label{suiso}
\end{equation}
We have $Q^aQ^b\eta_{ab}=J^2$, and $J$ is the total spin of 
the system.

Let us study the self-dual solitons
of the above system. The
Hamiltonian can be written as the Bogomol'nyi form as follows:
\begin{eqnarray}\label{sing}
H&=&\int d^2x \frac{2J^2\partial_i\psi^*\partial_i\psi}
{(1-\vert\psi\vert^2)^2}=
\int d^2x 
\frac{2J^2\vert(\partial_1\pm i\partial_2) 
\psi\vert^2}{(1-\vert\psi\vert^2)^2}
\mp 4\pi T,\nonumber\\
T&=&\frac{1}{4\pi}\int d^2x \left[iJ^2\epsilon_{ij}
\partial_i\left(\frac{\psi^*\partial_j\psi-\psi\partial_j\psi^*}
{1-\vert\psi\vert^2}\right)\right].
\end{eqnarray}
We find that the energy is saturated, when $\psi$ is
self-dual or anti-self dual, but these are singular solutions
because of  $\vert\psi\vert=1$. Both the energy and topological
charge diverge. We could restrict to the
upper sheeted hyperboloid given by $\vert\psi\vert<1$ or
the lower sheeted hyperboloid by $\vert\psi\vert>1$ from
the beginning, but nothing prevents
them from developing a divergence coming from the 
boundary $\vert\psi\vert=1$ in the integration.
We find that the CS gauge field enables to
circumvent this situation by providing an effective potential
in which the boundary  $\vert\psi \vert=1$ can not be reached
and make the energy and topological charge finite
(see  (\ref{effp}) and Fig. 1 in which the 
boundary  $\vert\psi\vert=1$ corresponds to $\phi=0$).

To see this, let us consider the gauged action of (\ref{xilag})
\begin{eqnarray}
S&=&\int dt d^2{x} \left[
iJ\frac{\psi^\dagger(D_0 \psi)-(D_0\psi)^\dagger\psi}
{1-\vert\psi\vert^2}-\frac{2J^2(D_i\psi)^\dagger D_i\psi}
{(1-\vert\psi\vert^2)^2}-\gamma_0 A_0-V(\psi)\right.
\nonumber\\
&&\left.+\frac{\kappa}{2}\epsilon^{\mu\nu\rho}
 A_\mu\partial_\nu A_\rho\right],
\label{lagra}
\end{eqnarray}
where $D_\mu = \partial_\mu -i A_\mu$, and $A_\mu$ is the CS
gauge field, and $\gamma_0$ is the constant background charge 
density.  We choose the potential energy to be
\begin{equation}
V(\psi)=\frac{\lambda}{2}(-Q^3+\gamma)(Q^3-v),
\label{potential}
\end{equation}
with $\gamma=\gamma_0+J$ and
$Q^3$ being given as in (\ref{suiso}). 
$\lambda$ is an anisotropic constant, and
$v$ is a free parameter connected with symmetry breaking~\cite{***}.  
This specific choice of (\ref{potential}) is for the purpose of the
self-duality. 
The Euler-Lagrange equation in terms of   
the non-compact spin variables $Q^a$ becomes
\begin{equation}
D_tQ-D_i[Q,D_iQ]+{\lambda}\Big(Q^3-\frac{v+\gamma}{2}\Big)
[Q,t^{3}]=0.
\label{eeq1}
\end{equation}
The gauge field equation is given by
\begin{equation}
\frac{\kappa}{2}\epsilon^{\mu\nu\rho}F_{\nu\rho}=j^{\mu}.
\label{eeq2}
\end{equation}
Here $j^{\mu}$ is a conserved current expressed by
\begin{equation}
j^{\mu}\equiv(\rho,j^{i})=(-Q^{3}+
\gamma,\;2\mbox{tr}(t^3[Q,D_{i}Q])),
\end{equation}
and then $Q_{{\rm U(1)}}=\int d^{2}x Q^3$ 
is the conserved $U(1)$ charge.

Let us first construct the Bogomol'nyi bound. Using the identity
\begin{eqnarray}
\frac{(D_i\psi)^\dagger D_i\psi}
{(1-\vert\psi\vert^2)^2}&=& \frac{
\left\vert
 (D_1\pm iD_2)\psi\right\vert^2}{(1-\vert\psi\vert^2)^2}
 \mp\frac{1}{2}\epsilon_{ij}F_{ij}
\left (\frac{1}{1-\vert\psi\vert^2}-v'\right)
\nonumber\\
&&\mp \frac{i}{2} \epsilon_{ij}\partial_i\left(
\frac{\psi^*\partial_j\psi-
\psi\partial_j\psi^*}{1-\vert\psi\vert^2} \right)
\pm\epsilon_{ij}\partial_i\left(\Big(\frac{1}{1-\vert\psi\vert^2}
-v'\Big)A_j\right),
\label{ident}
\end{eqnarray}
where $v'$ is an arbitrary constant, we
obtain the following Bogomol'nyi equation:
\begin{eqnarray}
H&=&\int d^2x\left(
\frac{ 2J^2\left\vert D_i{\psi}\right\vert^2}{(1-\vert\psi\vert^2)^2} 
+ V(\psi)\right)\nonumber\\
 &=&\int d^2x\left[ \frac{2J^2\left\vert(D_1\pm iD_2)
 \psi\right\vert^2}{(1-\vert\psi\vert^2)^2}
+V(\psi)\mp\frac{J}{2}\epsilon_{ij}F_{ij}
\left (Q^3-v^{''}\right)
\right] \mp 4\pi T,
\label{hamil}
\end{eqnarray}
where the topological charge is given by
\begin{equation}
4\pi T=\int d^2x\left[
 iJ^2 \epsilon_{ij}\partial_i\left(
\frac{\psi^*\partial_j\psi-\psi\partial_j\psi^*}
{1-\vert\psi\vert^2} \right)
-J\epsilon_{ij}\partial_i\left((Q^3-v^{''})
A_j\right)\right],
\label{topcharge}
\end{equation}
with  $v''=(2v'-1)J$.

Using the Gauss Law constraint
\begin{equation}
\frac{\kappa}{2}\epsilon_{ij} F_{ij}=
-Q^3+\gamma,
\label{gauss}
\end{equation}
and from (\ref{potential}) and (\ref{hamil}),
we find that self-dual limit is achieved 
with the condition $\lambda=\pm \frac{2J}{\kappa}$
and $v=v''(=(2v^\prime-1)J)$. Note that the latter condition
connects the symmetry breaking parameter with the 
topological charge. The Bogomol'nyi
bound is saturated by the topological charge
when the self-duality equation is satisfied:
\begin{equation}
\left( D_1\pm iD_2\right)\psi=0.
\label{123}
\end{equation}
Expressing the gauge field 
$A_i$ in terms of $\psi=\vert\psi\vert e^{i\Theta}$, we find
\begin{equation}\label{a}
A_j=\partial_j\Theta\pm\epsilon_{jk}\partial_k\ln\vert\psi\vert.
\end{equation}
Substitution into Gauss's law (\ref{gauss}) yields
\begin{equation}
\nabla^{2}\ln\vert\psi\vert\pm\epsilon^{ij}\partial_{i}
\partial_{j}\Theta
 =\pm\frac{1}{\kappa}\
\left( J\frac{1+\vert\psi\vert^2}{1-\vert\psi\vert^2}-\gamma\right).
\label{b}
\end{equation}

We concentrate on $J=1$ case for the time being.
It will be restored later when we discuss large $J$ limit.
Let us define $\vert\psi\vert\equiv e^{-\phi}$, 
then we obtain a scalar equation for the soliton configuration:
\begin{equation}
\nabla^{2}\phi\pm\epsilon^{ij}\partial_{i}\partial_{j}\Theta=
-\frac{dV_{eff}}{d\phi},
\label{pop}
\end{equation}
where the ``effective'' potential $V_{eff}$ is given by (See Fig. 1)
\begin{equation}\label{effp}
V_{eff}(\phi)=\pm\frac{1}{\kappa}(\ln\sinh\phi-\gamma\phi).
\end{equation}
Let us concentrate on the upper sign (self-dual  case) 
and the rotationally symmetric solutions.
The lower sign can be handled in a similar fashion. 
The ansatz in the
cylindrical coordinate  $(r,\theta)$ is given  by
\begin{eqnarray}\label{ansat}
\phi=\phi(r),\;\Theta=n\theta, \;
A_i=-\frac{\epsilon_{ij}x_j}{r^2}a(r).
\end{eqnarray}
Then, the Gauss's law (\ref{gauss})
and self-dual equation (\ref{123}) 
are given as
follows: 
\begin{eqnarray}
rf'(r)&=& [n-a(r)]\sinh f(r), \label{f'} \\
a'(r)&=&-\frac{r}{\kappa}[\cosh f(r)-\gamma].\label{a'}
\end{eqnarray}
Also the equation of motion in (\ref{pop})
becomes an analogue of the one
dimensional Newton's equation for $r>0$, if we regard $r$ as
``time'' and $\phi$ as the position of the hypothetical particle
with unit mass:
\begin{equation}\label{newtoneq}
\frac{d^{2}\phi}{dr^{2}}
+\frac{1}{r}\frac{d\phi}{dr}=\frac{1}{\kappa}\left(-\coth
\phi+\gamma\right).
\end{equation}
The exerting forces are the conservative force from the
effective potential $V_{eff}(\phi)$ in  (\ref{effp}),
and time-dependent friction $\frac{1}{r}\frac{d\phi}{dr}$.
When $n\ne 0$, there is an impact term at $r=0$ due to
$\epsilon^{ij}\partial_{i}\partial_{j}n\theta=\frac{n}{r}
\delta(r)$ in  (\ref{pop}).

The effective potential suggests that solutions are of three types;
the topological vortex with $n \neq 0~(\kappa >0,~ \gamma > 1)$, the
non-topological vortex with $n \neq 0 ~(\kappa <0,~ \gamma = 1)$, and the
non-topological soliton with $n=0~(\kappa<0,~ \gamma=1)$. 
In the first case, the
particle starts from $\phi=\infty$ at $r=0$ and moves until it reaches 
$\phi=\coth^{-1}\gamma$ at $r=\infty$. It corresponds to broken vacuum
with $Q^3=\gamma$.
In the second case, the particle starts from $\phi=\infty$,
reaches a turning point where it stops, changes the direction, and
finally rolls down again to $\phi=\infty$. In the last case, the
 particle starting at some finite position, rolls down directly to
$\phi=\infty$.
Let us examine the behavior of the solutions more closely.
Near $r=0$, the condition for $A_i$ to be non-singular forces $a(0)=0$
and $nf(0)=0$. Hence when $n\neq0$, $f(0)=0$
and when $n=0$, $f(0)$ is arbitrary. The behaviors of the solutions
near $r=\infty$ are determined by the 
condition $f'(\infty)=a'(\infty)=0$.\\
\noindent I. Topological vortex ($\kappa >0,~\gamma>1$).\\
Near $r=0$, trying power solutions of the form $f(r)= f_0r^p,
~a(r)= a_0r^q$,
we find $p=n$(positive), $q=2$, and $a_0=\frac{\gamma-1}{2\kappa}$.
(\ref{f'}) forces  $f'(r)>0$ near $r=0$. Hence $a(r)$ and
$f(r)$ are monotonically increasing function near $r=0$.
Near  $r=\infty$, since $\phi(\infty)=\coth^{-1}\gamma$, $f(\infty)$ is
$\ln(\gamma+\sqrt{\gamma^2-1})$
and the finiteness of energy requires $a(\infty)=n$.\\
\noindent II. Non-topological vortex ($\kappa <0,~\gamma=1,~n>0$).\\
Near $r=0$, we find $f= f_0 r^n$ and
$a(r)=-\frac{f_0^2}{4(n+1)\kappa}r^{2n+2}$ which are all
monotonically increasing functions near $r=0$. Near $r=\infty$, 
let us try
$f(r)=f_{\infty}r^{-p},~a(r)=\alpha+a_{\infty}r^{-q}$ where 
$\alpha\equiv a(\infty)$. Then we find  
$p=-n+\alpha,~q=2(\alpha -n-1)$, hence
$\alpha>n+1$.\\
\noindent III. Non-topological soliton ($\kappa<0,\gamma=1,~n=0$).\\
Near $r=0$, let us try $f(r)= f(0)+f_0r^p$ and $a(r)= a_0
r^q$. We find $p=q=2$ and $f_0>0$. 
This means that no solution exists in which the  ``particle''
climbs up the hill at first. It  rolls down from the beginning.
Hence $f(r)$ is a monotonically decreasing
function and $a(r)$ is an increasing function. Near $r=\infty$, let us
try $f(r)= f_{\infty} r^{-p},~a(r)= \alpha+a_{\infty}r^{-q}$.
We find $p=\alpha,~q=2(\alpha-1)$ and $a_{\infty}<0$,
hence $\alpha>1$.\\
Some numerical results are presented in Fig. 2 and 3.

Let us compute physical quantities of these solitons 
under the ansatz in (\ref{ansat}).
The topological charge (\ref{topcharge}) can be expressed by
\begin{equation}
T=\frac{n}{2}(\cosh f(0)-\cosh f(\infty))
-\frac{\alpha}{2}(\cosh f(\infty)-v^{''}).
\label{topcha}
\end{equation}
For topological vortex, we find $T=\frac{n}{2}(1-2\gamma+v^{''})$
and for nontopological vortex and soliton,
$T=-\frac{\alpha}{2}(1-v^{''})$.
The magnetic flux $\Phi\equiv\int d^2x \;B$ is given by
\begin{equation}
\Phi=2\pi\alpha.
\label{flux}
\end{equation}
It is $2\pi n$ for the topological vortex, and $2\pi(n+\beta)$,
with $\beta=\alpha -n$, for others.
We define the angular momentum~\cite{angular}
\begin{equation}\label{angm}
{M}=\int d^2x\; \epsilon_{ij}x_i\left\{i\frac
{(\partial_j \psi^*) \psi- \psi^*(\partial_j\psi)}
{1-\vert\psi\vert^2}
-A_j(Q^3-1)\right\}.
\end{equation}
We count only the unambiguous contribution coming
from the gauge field in the absence of 
the background charge $\gamma_0$.
It is given by
\begin{equation}\label{j}
{M}=\pi\kappa \alpha(\alpha-2n).
\end{equation}
${M}$ is in general fractional, 
which means the solitons are anyons. It tells
that ${M}=-\pi\kappa n^2$ for topological vortex,
${M}=\pi\kappa(\beta^2-n^2)$ for nontopological vortex, 
and ${M}=\pi\kappa\beta^2$ for nontopological soliton.

Next, we examine the large $J$ limit. To do that, let us consider 
the Holstein-Primakoff transformation and introduce $u$ defined by
\begin{equation}\label{hol}
u=\frac{\sqrt{2J}\vert\psi\vert}{\sqrt{1-\vert\psi\vert^2}}.
\end{equation}
Substitution of the inverse transformation
\begin{equation}
\vert\psi\vert=\frac{u}{\sqrt{2J+u^2}}
\end{equation}
into (\ref{b}) yields
\begin{equation}
\nabla^2\ln u -\frac12\nabla^2\ln(2J+u^2)=\pm \frac{1}{\kappa}
\left( u^2-\gamma+J \right).
\end{equation}
We find that in the limit $J\rightarrow\infty$ and $\gamma=J$, 
neglecting terms of order $1/J$, the
equation is precisely the Liouville equation of
the gauged nonlinear Schr\"odinger model \cite{jack-pi}. 
Also when $-\gamma+J=1$ it reduces to the Abelian Higgs
model~\cite{wein}.

\vspace{8mm}
\section{Quantization}
In this section, we consider quantization of the gauged spin system.
Since the Lagrangian of our model is singular, we use the Dirac's
constraint analysis~\cite{dirac}, together with 
the first-order method advocated by Faddeev and Jackiw \cite{fadd}.
We first represent the action (\ref{lagra}) 
in terms of the canonical variables
$(f, \Theta)$ given by $\psi=\tanh\frac{f}{2}e^{i\Theta}$.
To do that, we use  the non-compact spin variables;
\begin{equation}
Q^1=J\sinh f({\bf r})\cos\Theta({\bf r}),~
Q^2=J\sinh f({\bf r})\sin\Theta({\bf r}),~ 
Q^3=J\cosh f({\bf r}).
\label{new}
\end{equation}
We  denote vectors with bold face from here on.
The first order Lagrangian can be rewritten as
\cite{fadd}
\begin{equation}
L=  a_I({\xi})\dot\xi^I -H(\xi)\label{forder},
\end{equation}
where $\xi^I$ denotes collectively the canonical variables,
$( f({\bf r}),~\Theta({\bf r}),~A_i({\bf r}))$.
The time dependence will be omitted unless
needed. The Hamiltonian (\ref{hamil}) with the Gauss's law
constraint is given by 
\begin{equation}
H= \int d^2 {\bf r}\left[\frac{J^2}{2} 
\left(
(\partial_i f({\bf r}))^2+\sinh^2\!\! f({\bf r})\;
 (\partial_i\Theta-A_i)^2 \right)
+V+A^0 \tilde{G}
\right].
\end{equation}
The potential $V$ is given in (\ref{potential}) and
the Gauss's law is implemented as the first class
constraint;
\begin{equation}
\tilde{G} = \frac{\kappa}{2}\epsilon_{ij}F_{ij}+Q_3-\gamma\approx 0.
\label{gaus}
\end{equation}

The Euler-Lagrangian equation from (\ref{forder}) reads as
\begin{equation}
\Omega_{IJ}(\xi)\dot\xi^J = \frac{\partial H(\xi)}{\partial \xi^I},
\end{equation}
with $\Omega_{IJ}(\xi)=\frac{\partial}{\partial \xi^I}
a_J(\xi)-\frac{\partial}{\partial \xi^J}
a_I(\xi)$. $\Omega_{IJ}(\xi)$ defines the 
pre-symplectic two form $\Omega$ by
\begin{equation}
\Omega=\frac{1}{2}\Omega_{IJ}(\xi)d\xi^I d\xi^J = d a(\xi),
\end{equation}
where $a(\xi)$ is the canonical one form, $a(\xi) = a_I(\xi)d\xi^I$.
For the given first order Lagrangian, we find
\begin{equation} 
\Omega= \int d^2{\bf r}\left(J \sinh f({\bf r}) d f({\bf r})
d\Theta({\bf r})+\frac{\kappa}{2}
\epsilon^{ij} d A_i({\bf r}) d A_j ({\bf r})\right). \label{pre}
\end{equation}
When the matrix $(\Omega_{IJ})$ is non-singular, i.e., has 
its inverse $(\Omega^{IJ})$ as
in (\ref{pre}), the Poisson bracket is defined as
\begin{eqnarray}
&&\{F(\xi), G(\xi)\}= 
\Omega^{IJ}(\xi)\frac{\partial F(\xi)}{\partial \xi^I}
\frac{\partial G(\xi)}{\partial \xi^J}
\\
&&= \!\int\!\! d^2\!{\bf r}\left[ \frac{1}{J \sinh f({\bf r})}\!\!
\left(\frac{\partial F}{\partial f({\bf r})}\frac{\partial G}
{\partial \Theta({\bf r})}-\frac{\partial F}{\partial\Theta({\bf r})}
\frac{\partial G}{\partial f({\bf r})}\right)\!\!+
 \!\left(\frac{1}{\kappa}\epsilon_{ij}
\frac{\partial F}{\partial A_i({\bf r})}
\frac{\partial G}{\partial A_j({\bf r})}\right)\!\right],\nonumber
\end{eqnarray}
and the Euler-Lagrangian equations become 
\begin{equation}
\dot \xi^I = \{ \xi^I, H\} = 
\Omega^{IJ} \frac{\partial H}{\partial \xi^J}.
\label{ham}
\end{equation}
The above Poisson bracket yields
\begin{eqnarray}
\{J\cosh f({\bf r}),\Theta({\bf r'})\}&=
&\delta({\bf r}-{\bf r'}),\label{can}\\
\{A_i({\bf r}), A_j({\bf r'})\}& =& {1\over \kappa}\epsilon_{ij}
\delta({\bf r}- {\bf r'}).
\end{eqnarray}
Then, by using the expression (\ref{new}), we recover the 
$su(1,1)$ spin algebra
\begin{equation}
\{Q^a ({\bf r}),Q^b({\bf r'})\}=
-{\epsilon^{ab}}_{c}Q^c({\bf r})
\delta({\bf r}-{\bf r'}).
\label{spinning}
\end{equation}
One can check that no further secondary constraints arise
\begin{equation}
\{H, \tilde{G}\}\approx 0, 
\end{equation}
and  (\ref{ham}) yields the equations of motions
(\ref{eeq1}) and (\ref{eeq2}).

To quantize the system, we adopt the scheme which has
two properties. The first one is 
that we pursue the quantization
of dynamical variables which are functions of only $Q^a$'s,
which satisfy the basic $su(1,1)$ algebra (\ref{spinning}). 
This is due to the fact that not every observables which are
functions of the canonical variables satisfying
(\ref{can}) are quantizable \cite{geom}. Therefore, we prepare the
classical Hamiltonian in the form;
\begin{equation}
H= \int d^2 {\bf r}\left[\frac{1}{2}\left(
D_i Q^1 \!\!\cdot \!\!D_iQ^1+
D_iQ^2 \!\!\cdot \!\!D_iQ^2
-D_iQ^3 \!\!\cdot\!\! D_iQ^3\right)+
V(Q^3)+A^0 \tilde{G}\right].
\end{equation}
Then, we introduce
$Q_\pm=Q^1 \pm i Q^2,~ Q_0=Q^3$ and 
use the canonical relations coming 
from quantizing the classical algebra (\ref{spinning});
\begin{equation}
[Q_{+}({\bf r}),Q_{-}({\bf r}')]=
-2Q_{0}({\bf r})\delta({\bf r-r'}),~
[Q_{0}({\bf r}),Q_{\pm}({{\bf r}'})]=
\pm Q_{\pm}({\bf r})\delta({\bf r-r'}). 
\label{alge}\end{equation}
Secondly, we work in the reduced phase space 
quantization scheme \cite{jack-pi} in which the Gauss's law
is first solved to yield the quantum gauge potential 
$(\nabla=\frac{\partial}{\partial {\bf r}})$
\begin{equation}
{\bf{A}}({\bf r})=
{\bf \nabla}\times\frac{1}{\kappa}\int d^2{\bf{r'}}~~
G({{\bf r}-\bf{r'}})
[-Q_0({\bf{r'}})+\gamma],
\end{equation}
where $G({{\bf r}-\bf{r'}})$ is the Green's function;
\begin{equation}
G({{\bf r}-\bf{r'}})=\frac{1}{2\pi}
\ln\vert{{\bf r}-\bf{r'}}\vert.
\end{equation}
We will take ${\bf A(r)}$ to commute with $Q_{\pm}({\bf r})$,
$ Q_{0}({\bf r})$~\cite{jack-pi}.

Then, the normal ordered Hamiltonian is given by
\begin{equation}
H=\int d^2 {\bf r}
\;\left(\frac{1}{2}{\bf \Pi}^\dagger \cdot {\bf\Pi} -
\frac{1}{2}\nabla Q_0\cdot \nabla Q_0 + 
\frac{\lambda}{2}(-Q_0+\gamma)(Q_0-v)+A^0\tilde{G}\right),
\label{noham}
\end{equation}
where
\begin{equation}
{\bf{\Pi}}={\bf D}Q_-\equiv {\bf \nabla} Q_--i{\bf{A}} Q_-.
\end{equation}
The equation of motion is given by
\begin{eqnarray}
i\dot{Q}_-({\bf r}) 
& = & [Q_-({\bf r}),H] \nonumber \\
& = & -Q_{0}({\bf r})D^2Q_-({\bf r})+
\frac{1}{2}[Q_-({\bf r}),~{\bf \nabla}^2Q_{0}({\bf r})]_+
-A^0({\bf r})Q_{-}({\bf r})\nonumber\\
&&+\frac{1}{2\kappa^2}\int d^2{\bf{r'}}
{\bf\nabla} G({\bf r-r'})\cdot{\bf \nabla} G({\bf r-r'})
Q_{+}({\bf{r'}})Q_-({\bf{r'}})Q_-({\bf r})\nonumber\\
&&-\frac{\lambda}{2}\left([Q_-({\bf r}),~Q_{0}({\bf r})]_+
-(\gamma+v)Q_{-}({\bf r})\right),
\label{equmotion}
\end{eqnarray}
where $[~,~]_+$ denotes anticommutator.
The scalar potential $A^0$  is given by
\begin{equation}
A^0({\bf r})=-\frac{1}{\kappa}
\int d^2{\bf{r'}}~~G({\bf r-r'})
{\bf \nabla} \times {\bf{j}}({\bf{r'}}),
\end{equation}
where ${\bf{j}}$ is the current-density operator
\begin{equation}
{\bf{j}}=\frac{1}{2i}[Q_{+}({\bf r})
{\bf\Pi}({\bf r})-{\bf\Pi}^\dagger({\bf r})
Q_-({\bf r})].
\end{equation}
The third component $Q_0$ satisfies the current conservation
equation:
\begin{equation}
\dot{Q}_0+{\bf\nabla}\cdot{\bf{j}}=0.
\end{equation}

To give anyon interpretation, let us define a new operator
\begin{equation}
\hat{Q}_-({\bf r})=e^{-i\omega({\bf r})}Q_-({\bf r})
\equiv \exp \left[\frac{i}{2\pi\kappa}\int d^2{\bf{r'}}
\Sigma({\bf r-r'})(-Q_0({\bf r'})+\gamma)\right]Q_-({\bf r}),
\end{equation}
where $\Sigma$ is given by
\begin{equation}
{\bf\nabla}\times G({\bf r-r'})=
-\frac{1}{2\pi}{\bf\nabla}\Sigma({\bf r-r'}),~
\tan\Sigma({\bf r-r'})=\frac{y-y'}{x-x'}.
\end{equation}
We also define $\hat{Q}_0({\bf r})=Q_0({\bf r})$.
A straightforward computation shows that
the new operators satisfy the anomalous commutator
\begin{eqnarray}
\hat{Q}_{+}({\bf r})\hat{Q}({\bf{r'}})&=&
e^{\frac{i}{2\kappa}}
[\hat{Q}_{-}({\bf{r'}})\hat{Q}_{+}({\bf r})
-2\hat{Q}_0({\bf r})\delta({\bf r-r'})], \\\nonumber 
\hat{Q}_{\pm}({\bf r})\hat{Q}_{\pm}({\bf{r'}})&=&
e^{-\frac{i}{2\kappa}}
\hat{Q}_{\pm}({\bf{r'}})\hat{Q}_\pm({\bf r}), \\\nonumber 
 [ \hat{Q}_0({\bf r}),\hat{Q}_{\pm}({\bf{r'}}) ]&=&
\pm\hat{Q}_{\pm}({\bf r})\delta({\bf r-r'}).
\end{eqnarray}
Using the identity \cite{jack-pi}
\begin{equation}
-i[\omega({\bf r}),\partial_t\omega({\bf r})]=
\frac{1}{4\pi^2\kappa^2}\int d^2 {\bf{r'}}~~
{\bf\nabla} \Sigma({\bf r-r'})\cdot 
{\bf\nabla} \Sigma({\bf r-r'})
Q_{+}({\bf{r'}})Q_{-}({\bf{r'}}),
\end{equation} 
one can show that (\ref{equmotion}) becomes
 ``free'' equation in terms of $\hat Q$;
\begin{eqnarray}\label{free}
i\dot{\hat Q}_-({\bf r})&=&-{\hat Q}_0({\bf r})
{\bf\nabla}^2{\hat Q}_-({\bf r})+\frac{1}{2}
[{\hat Q}_-({\bf r}),\nabla^2{\hat Q}_0({\bf r})]_+\\\nonumber
&&-\frac{\lambda}{2}
\left([\hat{Q}_-({\bf r}),~\hat{Q}_{0}
({\bf r})]_+-(\gamma+v)\hat{Q}_{-}({\bf r})\right).
\end{eqnarray}
Hence, the our system provides another field theoretical
model of anyons \cite{han}: the gauge potential can be eliminated 
by a singular gauge transformation, and the multivalued
operators $\hat Q$'s satisfy a  free equation but anomalous
commutation relations.

To consider the large spin limits, let us introduce the 
Holstein-Primakoff transformation:
\begin{eqnarray}
Q_{-}({\bf r})&=&\sqrt{2J+\psi^\dagger({\bf r})
\psi({\bf r})}\psi({\bf r}),~~Q_{+}({\bf r})=
\psi^\dagger({\bf r})\sqrt{2J+
\psi^\dagger({\bf r})\psi({\bf r})},\nonumber\\
Q_{0}({\bf r})&=&J+\psi^\dagger({\bf r})\psi({\bf r}).
\label{hplimit}
\end{eqnarray}
One can check that the algebra (\ref{alge}) is 
satisfied if one postulates the following commutator \cite{miss}:
\begin{equation}
[\psi({\bf r}),\psi^\dagger({\bf{r'}})]
=\delta({\bf r-r'}),~
[\psi({\bf r}),\psi({\bf{r'}})]=
[\psi^\dagger({\bf r}),\psi^\dagger({\bf{r'}})]=
0.
\end{equation}
In the large $J$ limit, 
let us rescale $\lambda\rightarrow 2J\lambda$, 
$A^0\rightarrow 2 J A^0$, 
and choose $\gamma=v=J$. Then, substituting (\ref{hplimit}) into
(\ref{noham}) and keeping only the linear terms in $J$, 
we obtain (after we rescale $H$ by $2JH$, and perform
normal ordering with respect to $\psi$)
\begin{equation}
H=\int d^2 {\bf r}
\left[\frac{1}{2}{\bf \Pi}^\dagger \cdot {\bf\Pi} -\frac{\lambda}{2}
(\psi^\dagger\psi)^2+
A^0(\frac{\kappa}{2}\epsilon_{ij}F_{ij}+\vert\psi\vert^2)\right],
\end{equation}
which is the Hamiltonian of the gauged non-linear 
Schr\"odinger model of Jackiw and Pi~\cite{jack-pi}.

\section{Conclusion}

We found that the gauged non-compact spin 
system admits various self-dual 
rotationally symmetric solitons. Depending on the background charge,
they can be classified as the topological vortices, 
the nontopological vortices, and 
the nontopological solitons.
The topological lump solutions do not exist because of the
non-compact nature of the target space.
These soliton equations have  hidden large spin limits which
can be  reached by the classical Holstein-Primakoff realization.
By a suitable choice of the background charges,
the system reduces to the Liouville equation whose 
exact multi-soliton solutions are known, or the Abelian Higgs model.
In the second quantum aspect,
we found that the quantization of the gauged 
non-compact spin system leads to the nonrelativistic 
field theory describing anyons, and 
its large spin limit  is also connected with the gauged
nonlinear Schr\"odinger model.
This provides yet another realization of the CS field 
producing the long-range interaction which is  responsible for the 
fractional statistics.  

There remain several issues which deserve further investigation.
The first one is related with avoiding the singularity $\vert\psi\vert=1$
to obtain finite energy solitons. We found
that CS gauge field makes this possible by providing
an effective potential $V_{eff}$ of (\ref{effp}). 
This corresponds to introducing
a scale into the flat potential. In this respect, there remains other
alternatives, especially inclusion
of the fourth-order spatial derivative Skyrmion term \cite{Fadniemi}.
This term does not permit any self-duality in 2+1 dimension, 
but the existence of the
finite energy solitons itself would be interesting
(the same statement also holds in the compact case).
The analytic behaviors  of the self-dual solitons deserve further study.
The existence of  the multivortex solutions 
(with multicenters)~\cite{yang} of  the 
nonlinear partial differential equations described by  
(\ref{pop}) and (\ref{effp}) for the
various background charges needs to be
analyzed in detail.

The extensions into ``higher'' cases remain as open problems. First, we
can consider higher non-compact $CP(N)$ case \cite{oh}. 
The pivotal thing is to check 
whether the self-dual equations admit solitons with 
finite positive semi-definite 
energy. Gauging the full $SU(1,1)$ group or $SU(2)$ in the compact case
is another direction to try. 
In these cases, the Gauss's law (\ref{gauss}) will be
replaced by the non-Abelian version. This can be attacked either in
holomorphic gauge \cite{oh941} or in axial gauge \cite{bak1}. 
It would be interesting
to look into the self-dual equations in these gauges to find out whether
they lead to any sensible soliton theory. 
The quantization could also be performed in
these gauges, but the non-compact case will require 
a more careful treatment because
of the non-compact nature of the gauge field. 
The relations with the relativistic
$SU(1,1)$ model of \cite{kori} should be 
studied also to see whether the present
non-relativistic self-dual solitons could be obtained 
by taking some suitable limit.

On the more physical side, it would be meaningful if the 
CS gauged spin
system could be realized in the parity broken system. Especially  
note that similar type of vortices are analyzed in the  
double-layer fractional quantum Hall effect,
which is described by the compact
$CP(1)$ nonlinear sigma model coupled with the 
CS gauge field and a Hopf term \cite{ichi}. 
Since the non-compact symmetry is also connected with the quantum Hall
effect \cite{prui}, the vortices, solitons and anyon realization of the
simplest non-compact symmetry presented in this paper could have some
physical application.

\vspace{1cm}
\noindent{\large\bf Acknowledgments} \\
We would like to thank Dr. K. Kimm for useful discussions
and R. Jackiw for carefully reading the manuscript.
This work is supported in part 
by the Korea Science and Engineering Foundation 
through the CTP at SNU and the project number
(95-0702-04-01-3), 
and  by the Ministry of Education through the
Research Institute for Basic Science  (BSRI/97-1419).

\newpage
\begin{figure}[tbh]
\begin{center}
\epsfig{figure=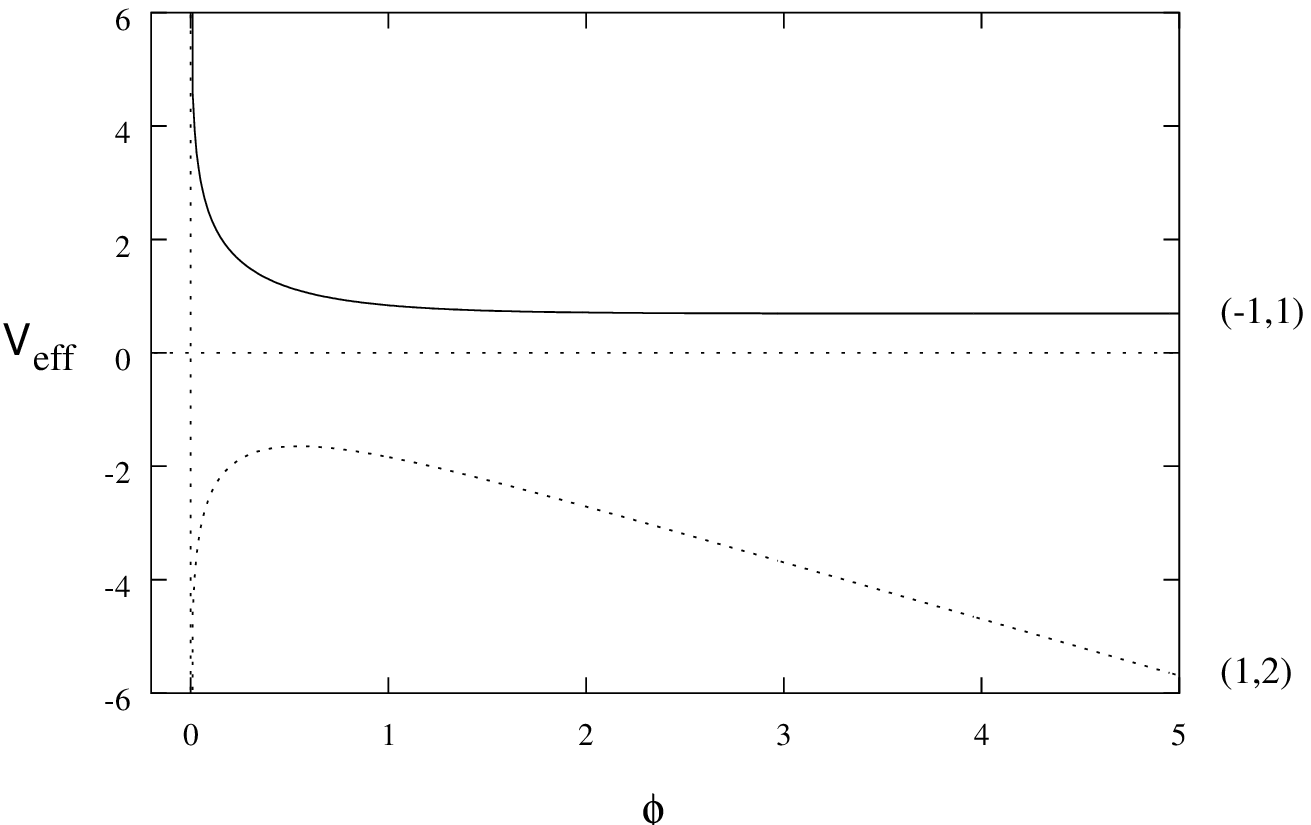,height=7cm}
\end{center}
Fig.1 Shapes of $V_{eff}(\phi)$ for some values of $\kappa$ and $\gamma$,
$e.g.,~(-1,1)$ means $\kappa=-1$ and $\gamma=1$.
\end{figure}
\newpage
\begin{figure}[tbh]
\begin{center}
\epsfig{figure=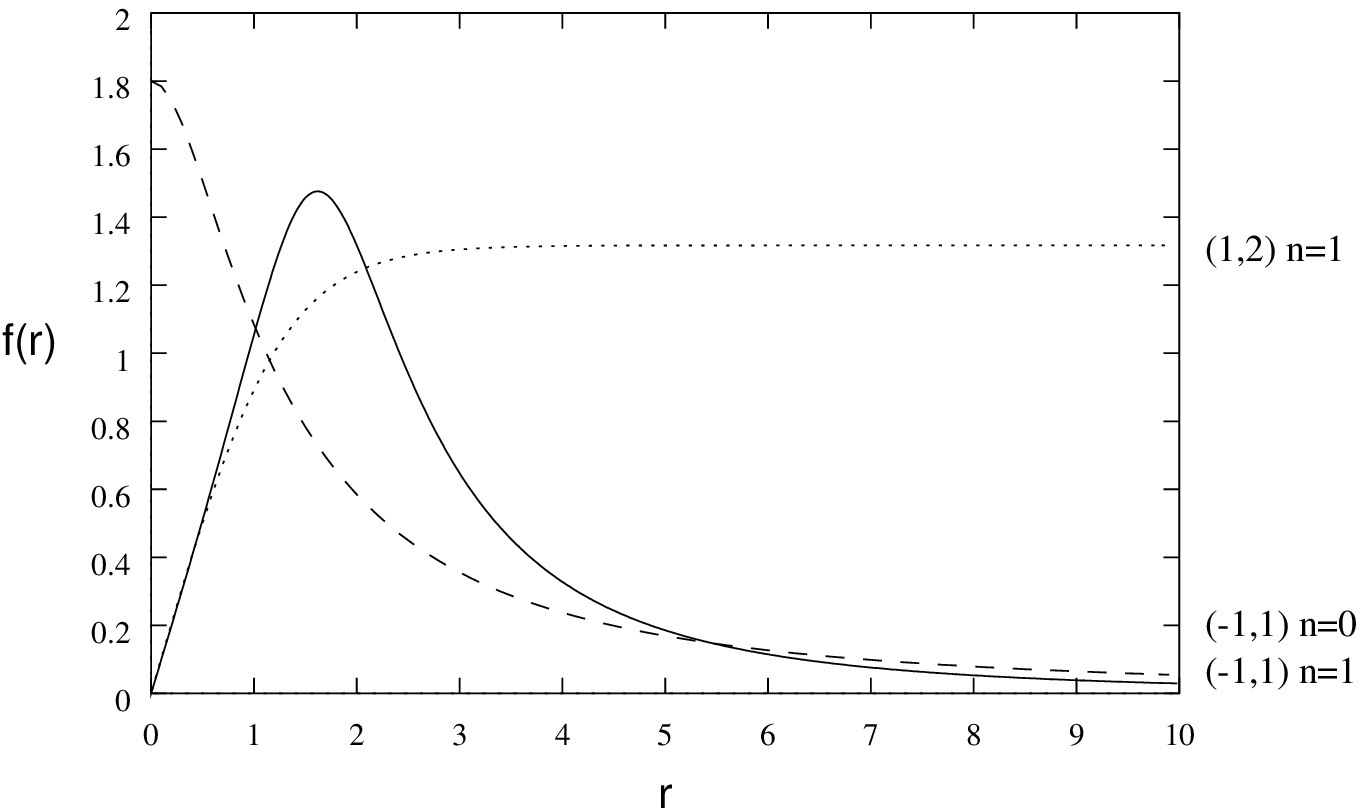,height=7cm}
\end{center}
Fig.2 Shapes of $f(r)$.\\ 
The solid line denotes non-topological vortex, the
dotted line topological vortex, and the dashed line
non-topological soliton. $(-1,1)$ means $\kappa=-1$
and $\gamma=1$.
\end{figure}
\newpage
\begin{figure}[tbh]
\begin{center}
\epsfig{figure=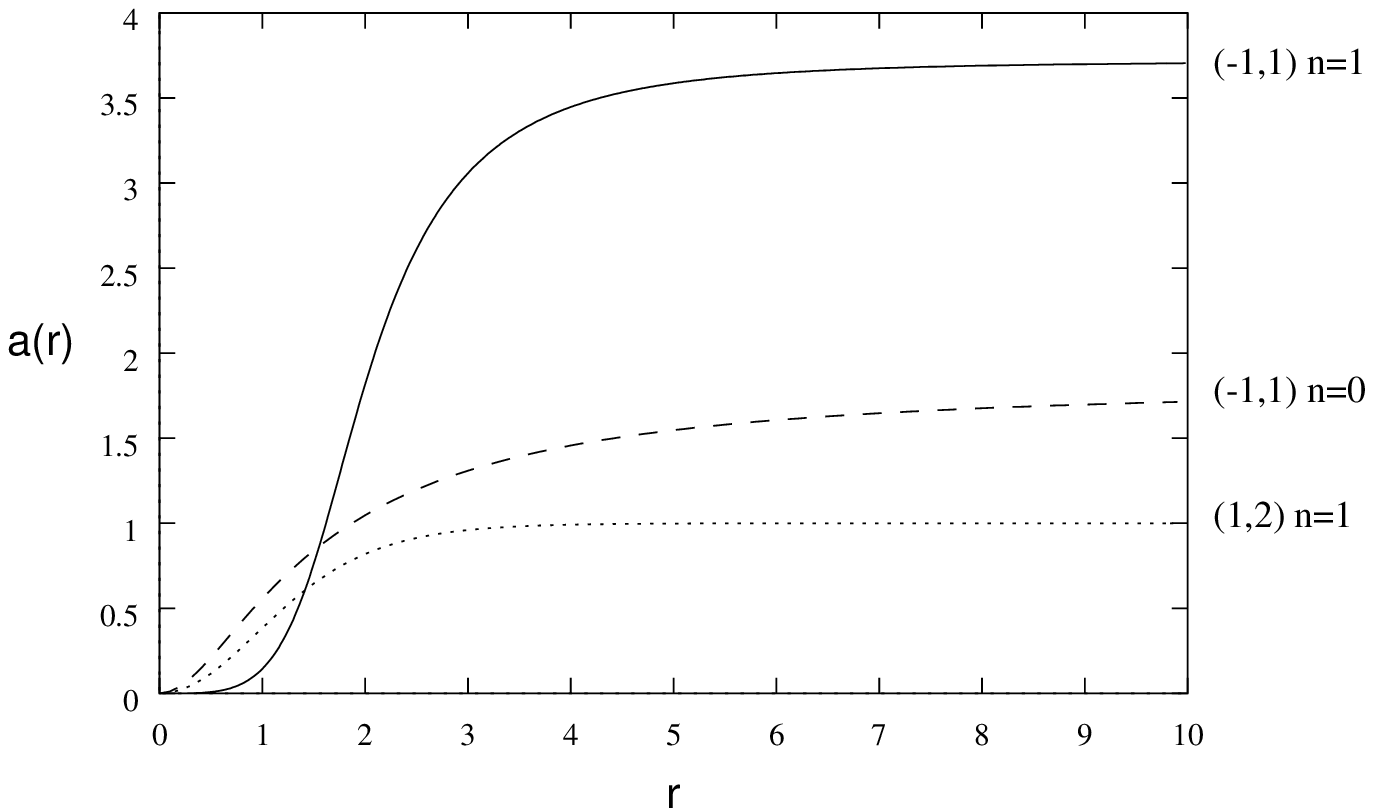,height=7cm}
\end{center}
Fig. 3 Shapes of $a(r)$.\\ 
The solid line denotes non-topological vortex, the
dotted line topological vortex, and the dashed line
non-topological soliton. $(-1,1)$ means $\kappa=-1$ 
and $\gamma=1$.                      
\end{figure}

\begin{thebibliography}{99}
\bibitem{miss} Note that in the compact case
with the spin algebra \\
$[Q_{+}({\bf r}),Q_{-}({\bf r}')]=
2Q_{0}({\bf r})\delta({\bf r-r'}),$~$
[Q_{0}({\bf r}),Q_{\pm}({{\bf r}'})]=
\pm Q_{\pm}({\bf r})\delta({\bf r-r'})$,\\ 
the Holstein-Primakoff realization is given by
 $Q_{+}({\bf r})=\sqrt{2J-\psi^\dagger({\bf r})
\psi({\bf r})}\psi({\bf r})$,\\
~$Q_{-}({\bf r})
=\psi^\dagger({\bf r})\sqrt{2J-\psi^\dagger({\bf r})\psi({\bf r})}
$,~$Q_{0}({\bf r})=J-\psi^\dagger({\bf r})\psi({\bf r})$
and we need opposite normal ordering.
\bibitem{oh} For example, the case of $SU(N,1)/SU(N)\times U(1)$ can 
be considered
which is described in  \cite{oh966}.
\end{thebibliography}
\end{document}